\documentstyle[11pt,paspconf]{article}
\input epsf

\def\lsim{\, \lower2truept\hbox{${< \atop\hbox{\raise4truept\hbox{$\sim$}}}$}\,}
\def\gsim{\, \lower2truept\hbox{${> \atop\hbox{\raise4truept\hbox{$\sim$}}}$}\,}
\def\ltsima{$\; \buildrel < \over \sim \;$}
\def\simlt{\lower.5ex\hbox{\ltsima}}
\def\gtsima{$\; \buildrel > \over \sim \;$}
\def\simgt{\lower.5ex\hbox{\gtsima}}
\def\deg{\ifmmode^\circ\else$^\circ$\fi}
\def\pdeg{\ifmmode $\setbox0=\hbox{$^{\circ}$}\rlap{\hskip.11\wd0 .}$^{\circ}
          \else \setbox0=\hbox{$^{\circ}$}\rlap{\hskip.11\wd0 .}$^{\circ}$\fi}

\newcommand{\Cel}{$^{\circ}$C}

\begin{document}

\title{CMB Synchrotron Foreground}
\author{George F. Smoot}
\affil{LBNL \& Physics Department, University of California, Berkeley CA 94720}

\begin{abstract}
Synchrotron emission is an important process in Galactic dynamics
and a potentially confusing foreground for cosmic microwave background (CMB)
radiation observations.
Though the mechanism of synchrotron emission is well understood,
the details for the Galaxy and many external sources is not well
characterized.
Quality maps at multiple frequencies are lacking but needed
for a full understanding of the Galactic synchrotron emission,
including intensity, spectrum, and spectral variation.
At high frequencies ($>$ 70 GHz) synchrotron emission
is not a severe limitation to precise CMB observations
well away from the Galactic plane.
\end{abstract}


\section{Introduction}

Synchrotron radiation is emitted by high energy electrons gyrating
in a magnetic field.
The first clearly understood observation came from the early betatron 
experiments in which electrons were first accelerated to relativistic energies
(Blewett 1946, Elder et al. 1947, 1948).
Although nonthermal radiation had been observed
from the Galaxy from the opening of radio astronomy in the pioneering work
by Karl Jansky in 1933,
there was no clear evidence of its origin.
In 1950 Kiepenheuer suggested that Galactic nonthermal radio emission was 
synchrotron radiation and Alfv\'en and Herlofson proposed that 
non-thermal discrete sources were emitting synchrotron radiation.
Kiepenheuer showed that the intensity of the nonthermal Galactic radio emission
can be understood as the radiation from relativistic cosmic ray electrons
that move in the general interstellar magnetic field.
He found that a field of $10^{-6}$~Gauss ($10^{-10}$ Tesla) and relativistic
electrons of energy $10^9$~eV would give about the observed intensity.
The early 1950s saw the development of these ideas 
(e.g. Ginzburg et al. 1951 and following papers, see Ginzburg 1969)
that synchrotron emission was the source of non-thermal ``cosmic'' radiation.
This model was later supported by maps which showed that
the sources of the non-thermal components were extended nebulae
and external galaxies
and by the discovery that the radiation was polarized
as predicted by theory.
The synchrotron theory is widely accepted and is the basis
of interpretation of all data relating to nonthermal radio emission.

In this model the observed non-thermal emission
is the emission of relativistic electrons in 
the weak Galactic magnetic fields.
The strength of the Galactic magnetic fields is, generally, 
such that only highly relativistic electrons can be responsible
for emission in the microwave and radio bands.

\begin{figure}
 \epsfxsize=5.7truein
\epsfbox{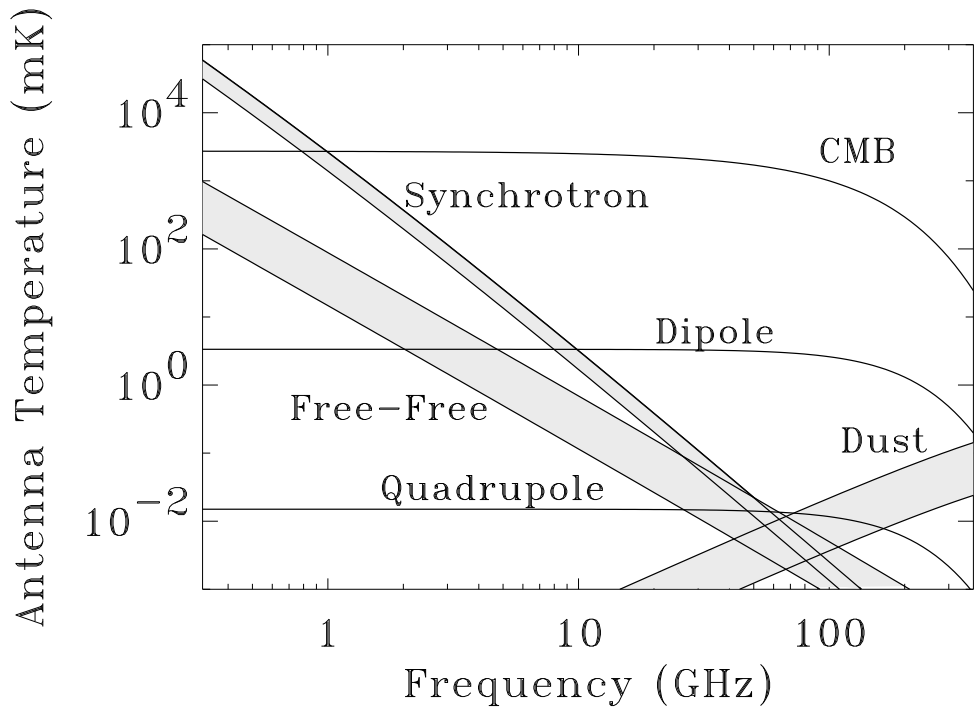}
\caption{
Graph showing frequency dependence and approximate relative strength
of Galactic synchrotron, free-free, and dust emission
as well as that of the cosmic microwave background and its features.
}
\label{FigGal}
\end{figure}

There are three diffuse Galactic emissions -- 
synchrotron, free-free, and dust --
which provide a significant confusing foreground to CMB anisotropy observations.
Figure \ref{FigGal} shows as a function of frequency
the approximate relative intensity of
the Galactic synchrotron, free-free, and dust emission
in relation to the cosmic microwave background (CMB).

The dust emission arise from the thermal reradiation of absorbed
stellar light.
The synchrotron emission is from relativistic cosmic-ray electrons originating 
in supernovae or other interstellar shocks and subsequently Larmor accelerated 
in interstellar magnetic fields. 
The free-free emission is the thermal bremsstrahlung from hot ($\gsim 10^4$K) 
%
%
electrons produced in the interstellar gas by the Galactic UV radiation field.

At meter wavelengths where synchrotron emission dominates the sky,
maps show a Galactic ridge about 5$^\circ$ wide and a much more
widely spread emission. 
The ridge, sometimes called the non-thermal disk, 
follows the Galactic plane and is
a few times broader than the hydrogen disk and thus 
wider than the Galactic plane free-free emission.
Extending well beyond this ridge is a very wide distribution of nonthermal
radiation, called the nonthermal halo or corona. 

Particularly 
at low Galactic latitudes the synchrotron and free-free emission 
are recognized as individual radio sources mostly 
associated with recent star formation.  
Thermal emission comes from HII regions in the vicinity of hot stars 
while the non-thermal synchrotron emission comes from supernova remnants.  
Radio surveys of the Galactic plane show in addition unresolved emission 
concentrated to the plane with the non-thermal emission being about twice 
the width of the thermal emission.  
Evidence for a wider distribution of low density electrons which result
in diffuse free-free comes from pulsar dispersion measure observations.  
These show the presence of a 1 kpc thick electron layer well 
beyond the denser 100 pc thick electron layer co-located 
with the OB star distribution.
Thus the synchrotron is wider than the free-free which in turn is wider
than the bulk matter distribution of the Galactic disk region.

Large features with a synchrotron spectrum extend far from the Galactic plane.  
The most prominent of these are the spurs and loops which 
describe small circles on the sky with diameters in the range 60$^\circ$ to 
120$^\circ$ (Berkhuijsen, Haslam \& Salter 1971).  
Because of their association with HI and in some cases with X-ray emission, 
they are believed to be low surface brightness counterparts 
of the brighter supernova remnants seen in lower latitude surveys 
such as that by Duncan et al. (1995) at 2.4 GHz. 
Other more diffuse structure at higher latitudes may be even older remnants.

\section{Astrophysical Synchrotron Emission Principles}

Synchrotron emission results from cosmic-ray electrons
accelerated in magnetic fields,
and thus depends on the energy spectrum of the electrons and 
the intensity of the magnetic field (Rybicki and Lightman 1979, Longair 1994).

\begin{figure}
\epsfxsize=5.0truein
\epsfbox{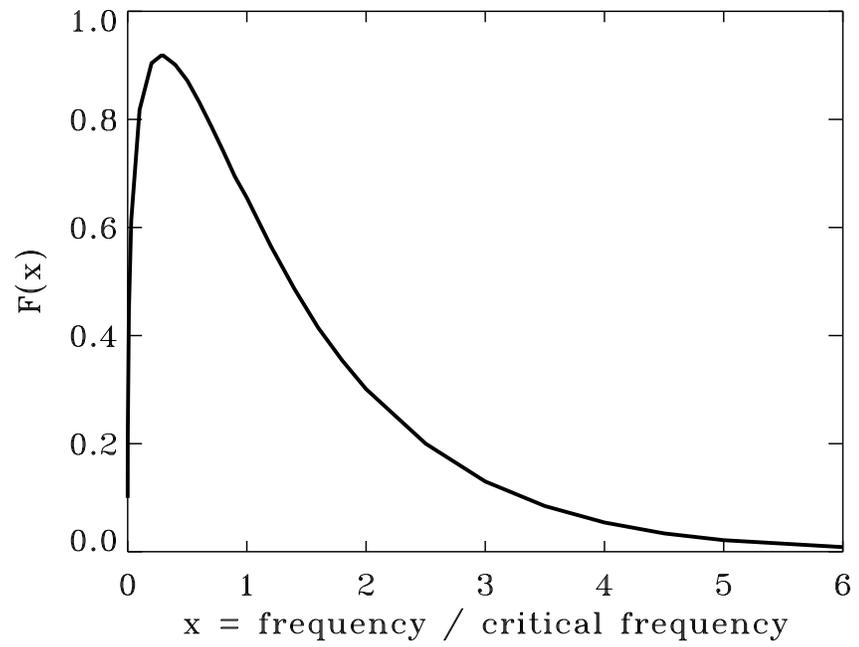}
\caption{
Graph showing intensity spectrum of single electron synchrotron radiation
plotted in terms of the dimensionless frequency
$x = \nu / \nu_c$, where $\nu_c$ is the critical frequency
$\nu_c = {3 \over 2} \gamma^2 {e B \over 2 \pi m_e} sin \alpha$
}
\label{FigSES}
\end{figure}

Electrons of a given energy ($E = \gamma m_e c^2$) 
radiate over a wide spectral band,
with the distribution peaking roughly at 
$\nu_c \approx 16.08 (B_{\rm eff}/{\rm \mu G })(E/{\rm GeV})^2\;{\rm MHz}$,
with a long low-power tail at higher frequencies,
and most of the radiation in a 2:1 band from peak.
Figure \ref{FigSES} shows this dependence.
The peak intensity is at
$\nu_{max} = 0.29 \nu_c 
= 4.6 (B_{\rm eff}/{\rm \mu G })(E/{\rm GeV})^2\;{\rm MHz}$.

The radiation from a single electron is elliptically polarized
with the electric vector maximum in the direction perpendicular
to the projection of the magnetic field on the plane of the sky.
Explicitly the total emissivity of a single electron via synchrotron
radiation is the sum of parallel and perpendicular polarization
\begin{equation}
j(\nu) = {\sqrt{3} e^3 B sin \alpha \over 16 \pi^2 \epsilon_0 c m_e} F(x)
\end{equation}
where $\alpha$ is the electron direction pitch angle to the magnetic field $B$
and $F(x) \equiv x \int_x^\infty K_{5/2}(\eta) d \eta $ 
is 
shown graphically in Figure \ref{FigSES}.

The quantity $x$ is the dimensionless frequency defined as
$x \equiv \omega / \omega_c = \nu / \nu_c$
where $\omega_c$ and $\nu_c$ are the critical synchrotron frequencies.
An electron accelerated by a magnetic field $B$ will radiate.
For nonrelativistic electrons the radiation is simple and called
cyclotron radiation and its emission frequency is simply
the frequency of gyration of the electron in the magnetic field.

However, for extreme relativistic ($\gamma \gg 1$) electrons the frequency
spectrum is much more complex and extends to many times the gyration frequency. 
This is given the name synchrotron radiation.
The cyclotron (or gyration) frequency $\omega_B$ is 
\begin{equation}
\omega_B = {q B \over \gamma m c}
\end{equation}
For the extreme relativistic case, aberration of angles cause the
radiation from the electron to be bunched and appear as a narrow pulse
confined to a time period much shorter than the gyration time.
The net result is an emission spectrum characterized by a critical frequency
\begin{equation}
\omega_c \equiv {3 \over 2}  \gamma^2 \omega_B sin \alpha
= {3 \gamma^2 q B \over 2 m c} sin \alpha
\end{equation}

To understand the astrophysical radiation, one must consider that
cosmic ray electrons are an ensemble of particles of different
pitch angles $\alpha$ and energies $E$.
It can generally be assumed that the directions are fairly isotropic
so that integration over pitch angles is straightforward.
The next step is integration over electron energy spectrum
to determine the total synchrotron radiation spectrum.

If the electrons' direction of motion is random with respect to the
magnetic field, and the electrons' energy spectrum can be approximated
as a power law: $ dN /dE = N_0 E^{-p}$, then the luminosity is given by 
\begin{equation}
I(\nu) = {\sqrt{3} e^3 \over 8 \pi m c^2} 
\left( {3 e \over 4 \pi m^3 c^5} \right)^{(p-1)/2}
L N_0 B^{(p+1)/2}_{\rm eff} \nu^{-(p-1)/2} a(p) ,  
\end{equation}
where $a(p)$ is a weak function of the electron energy spectrum (see Longair,
1994, vol.\ 2, page 262 for a tabulation of $a(p)$),
$L$ is the length along the line of sight through the emitting volume,
$B$ is the magnetic field strength, and $\nu$ is the frequency.

At very low frequencies synchrotron self-absorption is very important
as according to the principle of detailed balance, to every emission
process there is a corresponding absorption process.
At the lowest frequencies synchrotron self-absorption 
predicts an intensity that increases as $\propto \nu^{5/2}$.
This does not follow the Rayleigh-Jeans law because the
the effective kinetic temperature of the electrons varies with frequency
($T_e \approx (m_e c^2 / 3 k) (\nu/\nu_g)^{1/2}$ Longair 1994).
Spectra of roughly this form are found at radio, centimeter, and 
millimeter wavelengths in the nuclei of active galaxies
and quasars.
For our Galaxy the flux density increases at long wavelengths 
($\lambda > 30$~m) roughly proportional to $\sqrt{\lambda}$.
This is an effect that is due to self- and free-free absorption.

The local energy spectrum of the electrons has been measured 
to be a power law to good approximation, 
for the energy intervals describing the peak of radio synchrotron emission 
(at GeV energies).  
The index of the power law appears to increase from about
2.7 to 3.3 over this energy range (Webber 1983, Nishimura et al 1991).  
Such an increase of the electron energy spectrum slope is expected, 
as the energy loss mechanisms for electrons increases with 
 the square of the electron energy.

The synchrotron emission at frequency $\nu$ is dominated by cosmic ray
electrons of energy 
$E \approx 3({\nu/{\rm GHz}})^{1/2}\;{\rm GeV}$.
The range of energies contributing to the radiation intensity
at a given frequency depends on the electron energy spectrum:
the steeper the electron distribution, the narrower the energy range
(Longair 1994).
For the case of most of the Galaxy, this range is of order 15 to 50.
The observed steepening of the electrons' spectrum at GeV
energies is used to model the radio emission spectrum at GHz frequencies
(e.g. Banday \& Wolfendale, 1990, Platania et al. 1998).

\section{Low-Frequency Sky Surveys}

\begin{figure}
 \epsfxsize=5.3truein
 \epsfbox{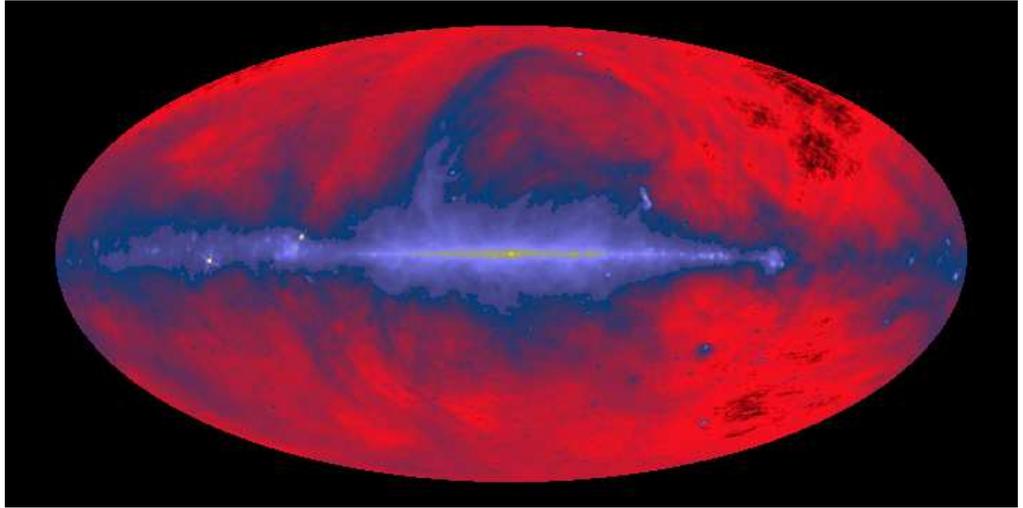}
\caption{
Color rendition of the 408 MHz map of Haslam et al. 
}
\label{Fig408}
\end{figure}

Over the years several researchers have attempted to map the antenna 
temperature of the sky at radio frequencies. 
Radio astronomers have concentrated on the study of discrete 
objects at high resolution, rather than on the investigation of the diffuse foreground. Radio 
maps of the Galaxy are not readily available. 
At present the only existing full-sky survey is the 408 MHz survey 
by Haslam et al. (1982) shown in the color figure \ref{Fig408}.
It consists of observations conducted with several instruments 
over a 10 year span. 
Although it has excellent angular resolution (0.85$^\circ$), 
the errors associated with the zero level 
and gain stability of the data are sufficiently large to produce significant 
uncertainties in the extrapolation of the measured sky brightness 
to higher frequencies, and, most seriously, 
there is no comparable quality map at another frequency for extrapolation
that accounts for spectral index variation.
Maps with partial sky coverage have been made at frequencies between 35 and 
1500 MHz (see Lawson et al. 1987 for a discussion). 
Table \ref{TabMaps} summarizes existing maps with significant northern 
hemisphere coverage. 
The survey by Reich et al. (1986) at 1420 MHz covers declination $d > 19^\circ$ 
(the Northern and Equatorial sky) and full sky coverage is planned. 
Incomplete coverage also exists at 38 (Milogradov-Turin, 1984), 
178 (Turtle \& Baldwin, 1962), 404 (Pauliny-Toth \& Shakeshaft, 1962), 
and 820 (Berkhuijsen, 1972) MHz. 

Another survey at 2300 MHz with nearly full coverage for $d < ~30^\circ$ is 
under way (Jonas et al. 1998). 
This new survey has good angular resolution (~0.9$^\circ$), but lacks 
a determination of the zero point of the temperature scale. 
It shows irregularities and patchy areas, 
resulting from the use of different receivers and feed antennas, 
as well as ``striping", 
which results from altitude-dependent ground emission contributing 
through the antenna sidelobes.

\begin{table}
\caption{
Summary of the radio maps available in the literature. Where the sky coverage 
is incomplete, the missing part lies always in the Southern Hemisphere. 
The one southern map at 2300 MHz from South Africa (Jonas et al., 1998) 
is not included here since its data are still unpublished. 
Maps of discrete sources (e.g. Condon \& Broderick, 1986, Condon et al. 1989) 
are not included because the diffuse 
continuum has been biased out. See text for references. }
\label{TabMaps}
\begin{center}\scriptsize
\begin{tabular}{ccccccc}
Frequency & Resolution & Gain & \multicolumn{2}{c}{Zero-level} & Zero-level & Sky \\
 & & uncertainty & \multicolumn{2}{c}{uncertainty} & correction\tablenotemark{a} & coverage \\
(MHz) & (deg) & (\%) & (\% of min)\tablenotemark{b} & (K) & (K) & (\%) \\
\tableline
38 & 7.5 & 5 & 4 & 300 & --- & 72 \\
178 & 0.22 x 5 & 10 & 15 & 15 & +27.5 & 57 \\
404 & 8.5 x 6.5 & 3.6 & 25 & 2.3 & N/A & 68\tablenotemark{c} \\
408 & 0.85 & 10 & 27 & 3 & --- & 100 \\
820 & 1.2 & 6 & 40 & 0.6 & --0.97 & 57 \\
1420 & 0.6 & 5 & 140 & 0.5& --0.13 & 68 \\
\end{tabular}
\end{center}

\tablenotetext{a}{Residuals of power-law fit to 38 \& 408 MHz surveys
of minimum sky brightness in the Northern Hemisphere; from Lawson et al. 1987.}
\tablenotetext{b}{Fractional error in the region of minimum emission after 
a 2.7 K correction for the CMB.}
\tablenotetext{c}{The sky was under sampled at the resolution of the survey.}

\end{table}

\begin{figure}
\epsfxsize=5.0truein
\epsfbox{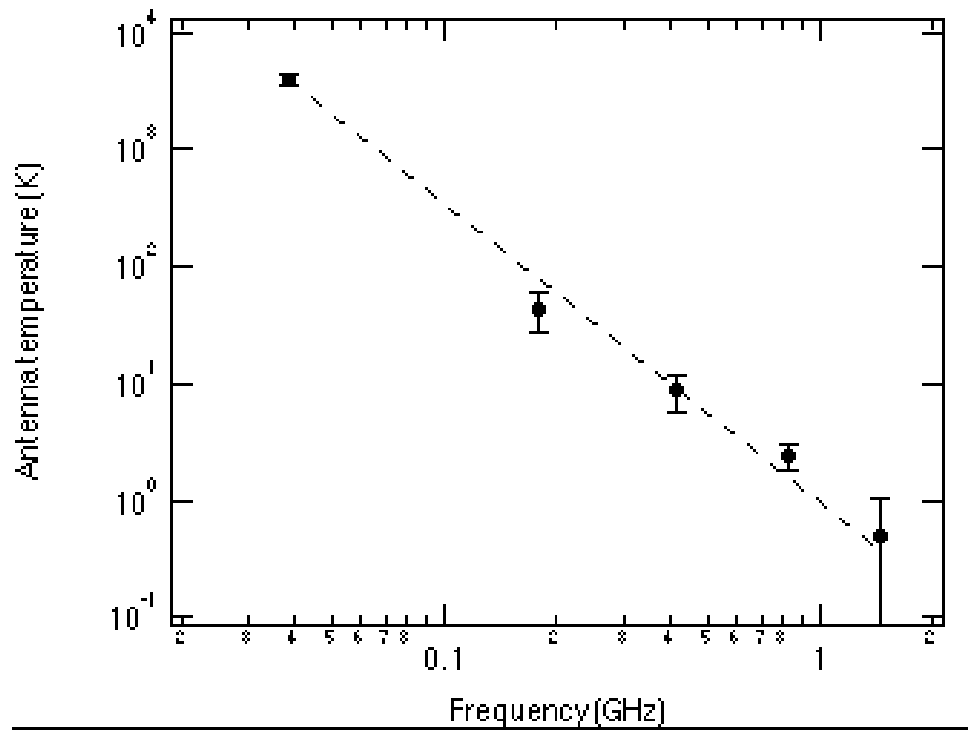}
\caption{
An extrapolation of the synchrotron spectrum using a power law model from 
measurements of the minimum of Galactic emission.
}
\label{FigSS}
\end{figure}

These maps have angular resolution ranging from excellent 
(0.8$^\circ$ at 1420 MHz) to adequate (7.5$^\circ$ at 38 MHz) 
but shortcomings arise when they are used to 
determine the spectral index and absolute level of sky radiation. 
In particular, large corrections are needed to reconcile the published surveys 
with the power-law dependence of $I(\nu)$ as described previously
(Lawson et al. 1987). 
The required corrections are comparable with the map errors quoted in Table 1. 
The relative internal gain uncertainty for the 408 
and 1420 MHz surveys is 10\% and 5\%, respectively.  
The  absolute  zero-level calibration,  
has an error of 3 K and 0.5 K, for the 408 and 1420 MHz surveys, respectively. 
In regions of low sky brightness these errors result in more 
than a $\pm 36$\% error in the calculated spectral index.
Even after introducing corrections to the zero level, 
the relative size of the error bars is so large that the 
spectral index is poorly determined for all frequency 
ranges (see Figure \ref{FigSS}). 

The internal self-consistency of the 1420 MHz map is so poor that the
measurements of the CMB at 1420 MHz have preferred to rely on convolutions of 
the 408 MHz map for Galactic subtraction (Staggs 1992, priv. comm.).
The task of determining the spectral index and its spatial variations 
over the sky is also hampered by the fact that the measurements presently 
available were made either: 

\noindent
1) with different instruments from different sites 
(opening the possibility of systematic errors 
associated with merging the different data sets as with the 408 MHz map), or 

\noindent
2) by using one instrument and tipping it to various zenith angles, 
which introduces a variable (and usually poorly measured) contribution 
from ground emission (as in the maps of Reich \& Reich and Jonas et al.), 
and often both problems coexist in the same data set.
Of particular interest for the CMB observations, 
especially for satellite missions, are surveys 
of significant areas of the sky at higher frequencies, e.g. 5-15 GHz.  
Only limited data close to the Galactic plane or in selected areas 
at higher latitudes with restricted angular resolution are available.  
Surveys of about one steradian of the sky have been made at 5 GHz 
by Melhuish et al. (1997) at Jodrell Bank and 
at 10, 15 and 33 GHz by Gutierrez et al. (1995) with the Tenerife experiments.

\subsection{THE GEM PROJECT}

Recognizing this problem of lack of good maps and observations
that could be used to extrapolate to higher frequencies in 1991 we initiated 
the Galactic Emission Mapping (GEM) program. 
The goals of GEM are to make maps overlapping with previous large sky
surveys and then have high resolution and sensitivity maps
at the higher frequencies of interest to CMB observations
and where the spectral index variations would become manifest.
The project paper (De Amici et al. 1994),
the 408 analysis paper (Torres et al. 1996) and technical notes
give details of the project, instrumentation and data analysis. 
See also the GEM project home page
http://aether.lbl.gov/www/projects/gem/
  
One of the major difficulties faced by a full sky radio survey 
is the need to achieve accurate calibration. 
The GEM design strategy consists of a `portable' radio-telescope 
that can be moved to sites at different latitudes. 
Using the same calibrated instrument allows 
a consistent merging of sections of the sky observed from different sites. 
An additional advantage is that moving the instrument at different latitudes 
allows pointing the main beam at a small angle from the local zenith, 
thus minimizing sidelobe contributions and the atmospheric absorption 
(important at the higher frequencies) seen at large zenith angles.

The GEM telescope is mounted on a rotating base and 
the pointing of the main beam is kept at a fixed angle from the zenith
to keep the mean atmospheric contribution constant.
The mechanical design of the mount system 
allows for changes in the zenith angle of the antenna. 
The antenna sidelobes can be checked by
observing the same portion of sky at different zenith angles,  
therefore obtaining a direct measurement of ground contribution.
The combined motion of the rotating base and the Earth's rotation results 
in a circle moved in a swath across the sky seen at fixed latitude. 
In principle one could cover the whole sky by collecting 
data in this mode from three equidistant latitudes 
(i.e. $60^{\circ}$S, $0^{\circ}$ and $60^{\circ}$N).
In practice it is desirable to allow for overlap
of the covered regions so as to ensure a self-consistent data set. 

GEM has acquired several hundred 
hours of observation at 408, 1465, 2300 and 5000 MHz
from an Equatorial site in Colombia and
a Northern site in Bishop, California. 
Fewer observations were made at the IAC-Tenerife Observatory in Spain. 
The GEM instrument was then moved to Caehoira Paulista, Brazil
for refurbishment and has begun new observations already at 1.4 and 2.3 GHz.
GEM is currently operating and being upgraded in Brazil udner 
the direct supervision of Thyrso Villela's group 
from INPE ${\rm Sa \tilde o}$ Jos\'{e} dos Campos with Camilo Tello 
handling on-site operations.

\subsubsection{GEM Hardware Design and Performance}

\subsubsection{Experimental Concept} 
\begin{figure}
 \epsfxsize=5.3truein
\epsfbox{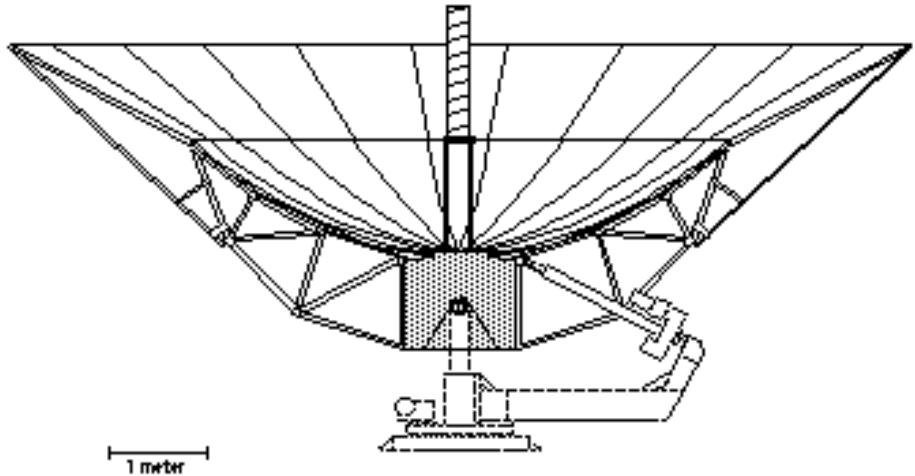}
\caption{
Schematic showing the basic layout of the GEM instrument.
}
\label{FigGEM}
\end{figure}

A block diagram (Figure \ref{FigGEM}) shows the main 
components that form the GEM system.  
The main section is made of the parabolic reflector, halo extension shield,
the feed antenna, the rotating base and the receiver. 
Table~\ref{tble:param} summarizes 
the instrumental parameters of GEM and two receivers parameters.

\begin{table} [htb]
\caption{GEM parameters} \label{tble:param}
\begin{center}
\begin{tabular}{ll} \hline
Parameter                &   value \\ \hline \hline
System \\
Receiver frequencies: & 0.408, 1.465, 2.3, 5 \& 10 GHz \\
Reflector diameter              &  5.5 m  \\ 
Reflector diameter with extension & 9.5 m \\
Base rotation speed (nominal)     & 1 rpm   \\
\hline
408 MHz Receiver         &   \\
System temperature       & $104 \pm 6$ K \\
Gain                     & $58 \pm 1$ K V$^{-1}$ \\
Band width               & 28 MHz \\
Sensitivity              & 26 mK/integration time \\
Beam width (FWHM)        & $11.3^{\circ}$ \\
Baseline susceptibility  & $-3$ K \Cel $^{-1}$ \\
Gain susceptibility  & $-2.7 \times 10^{-4}$ \% \Cel $^{-1}$ \\
\hline
2300 MHz Receiver parameters  &   \\
Wavelength (nominal)  & 13.04 cm \\
System temperature (nominal) & 55 K \\
System temperature (effective) & $66 \pm 11$ \\ 
Gain               & $40 \pm 7$  K V$^{-1}$ \\
Band width         & 100 MHz \\
Sensitivity        & 12 mK per sample \\
Beam width   & FWHM = $3.7^{\circ}$  \\
Main beam solid angle & $4.1 \times 10^{-3}$ \\
Directivity           & 3063 \\
Effective aperture    & 4.2 m$^2$  \\
Aperture efficiency   & 0.18 \\
Baseline susceptibility & $-0.039$ V \Cel $^{-1}$ \\
\hline
\end{tabular} 
\end{center}
\end{table}

GEM uses a Scientific Atlanta  5.5-m parabolic 
reflector mounted on an alt-azimuth rotating base. 
The 408, 1465 and 2300 MHz receivers use a prime focus feed: 
a backfire helix at the low frequencies, and a conical antenna at 2300 MHz.
The 5 and 10 GHz receiver units operate in differential mode with
two conical antennae. 
One is mounted at the Gregorian focus and the other points 
at the zenith.
Aluminum panels extend the parabolic reflector 
surface to a total diameter of 9.5-m. The purpose 
of this aluminum shield is to minimize diffracted
ground emission and to the beam efficiency and loss by
covering up with highly reflective opaque 
material half the $4\pi$ solid angle.

The 408 MHz and 1.465 GHz systems employ back-fire helix feed antennas, 
which consists of a 9.5 turns 
of length is $0.92 \lambda$.
The feed antennas are sensitive to circularly polarized radiation.
The main lobe width of the combined antenna/reflector assembly
is obtained using the transit of the Sun.

The mount assembly rests on a rotating base with rotation rate of about 1 rpm.
An azimuth angle encoder is mounted on the rotating axis of the main
GEM assembly and a similar encoder mouned on the horizontal axis 
to measure the altitude angle.
The zero angle resulting from this reading is calibrated 
using the Sun signal in the data in combination with the sun ephemerides.  
Calibration parameters were verified by checking GEM's mechanical orientation 
with respect to the geographic North employing a theodolite. 

We discuss two receivers as examples of the GEM system:

\subsubsection{The 408 MHz receiver}

The 408 MHz radiometer uses a total power receiver 
with two RF amplification stages and one DC amplification 
($\times 1000$) after detection. 
A cavity filter ($\Delta \nu = 28$ MHz) at the front end of the receiver and 
a tubular filter after RF amplification are used. 

The ambient temperature of the receiver is controlled and isolated 
from the outside temperature by warming the inside of the receiver box 
with resistance heaters while cooling the inside of a bigger box that
encloses the receiver box. 
Temperature sensors and a regulating circuit
keep the  operating temperature of the receiver to within $\pm 0.2$ \Cel.


Gain variations are monitored by injecting a fixed amplitude reference pulse 
every 45 seconds. 
The reference noise pulse is generated by a diode and is connected to the
input of the first amplifier stage via a directional coupler.

Radio frequency interference (RFI), due to local radio communications or 
electric discharges in the atmosphere can cause 
excessive dispersion of the observed signal.
An RFI detection circuit signals the presence of such anomalous data.
This circuit produces an output voltage $V_{\rm sat}$ proportional 
to the number of times that the signal average 
crosses some preset threshold during the integration time. 
Cutting data above some $V_{\rm sat}$ threshold effectively
acts as a low-pass filter on the signal. 

\subsubsection{2.3 GHz Receiver}

The 2.3 GHz receiver is mounted at the prime focus of the GEM telescope 
and fed by a cylindrical feed horn. 
The feed horn has a symmetric beam of $120 \deg$ at -12 dB
to properly illuminate the primary, and a measured VSWR of 1.023.

The receiver is a direct-gain total-power radiometer.
After the waveguide-coax transition,
a section of semirigid coax connects the first amplifier, 
an HEMT (High Electron Mobility Transistor)
amplifier (Berkshire Technologies S-2.3-30-RH) with flat 
response in the nominal 2.25 to 2.35 GHz band.
The total gain is about 35 dB and the noise temperature is 
$T_{amp}\approx 30$ K at an operating temperature of 300 K.
The signal is attenuated (2-4 dB) before the second amplifier
(MITEQ AFS4-02200240-20-10P-4), which provides a gain of 45 dB. 
After the second stage of amplification
a tubular pass band filter (TRILITHIC 4BC2300/100-3-KL)
with a 100 MHz pass band at 3dB rejects out-of-band noise.
Then, after a second attenuator,
a square-law diode detector rectifies the signal. 
The DC voltage is amplified by a factor of 500 (low gain) or 1000 (high gain),
digitized and integrated with an integration time of 0.56 s.
The overall $T_{sys}$ is about 55 K, while the total amplification is 
about 70 dB (including attenuation). 

The RF-chain is mounted on a aluminum plate, enclosed into a RF-tight box, 
and it is thermally controlled by temperature sensors and heaters. 
This box encloses also the throat section of the feed antenna and, on its
top, matches the GEM mounting interface through a latching mechanism.

\subsubsection{The GEM 2.3 GHz Sky Map}

We have chosen 2300 MHz as one of the frequencies for GEM so that we can 
complete the sky coverage of the Jonas' map to the northern hemisphere; 
large overlapping areas between the two maps will allow direct transfer 
of the absolute calibration to the southern map, 
thereby complementing their effort and increasing the scientific usefulness 
of the existing data.

COBE pixelization level 6 (ie. 6144 pixels of size $\sim 2.9^{\circ}$)
was used for the pixelized database, which seems appropriate
given the approximate beam width of $\sim 4^{\circ}$.

\begin{figure}
 \epsfxsize=5.3truein
 \epsfbox{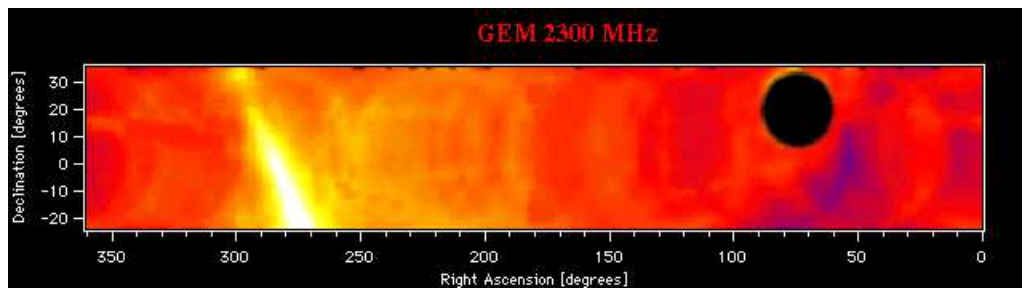}
\caption{
Color rendition of the GEM 2.3 GHz map made from Colombia. 
Note hole where Sun signal obscurred Galactic emission during data taking.
}
\label{FigGEM2CA}
\end{figure}

\begin{figure}
 \epsfxsize=5.3truein
 \epsfbox{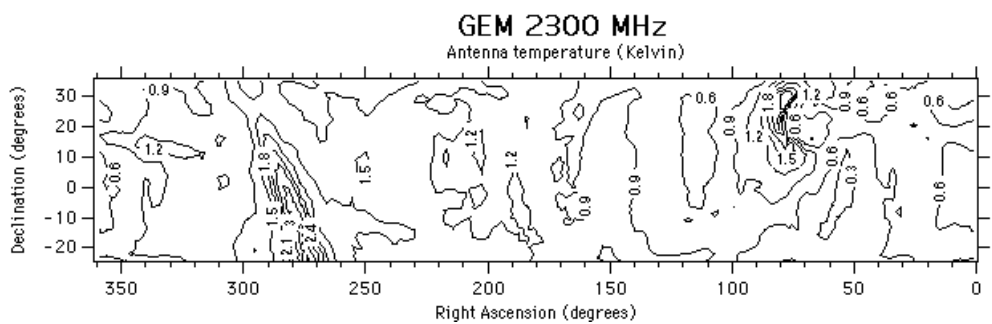}
\caption{
Contour plot of the GEM 2.3 GHz map made from Colombia.
}
\label{FigGEM2CB}
\end{figure}

The primary output of the analysis pipeline is 
a sky map with 3289 pixels out of 6144 implying 
a 53 \% sky coverage corresponding to the 
$60^{\circ}$ wide celestial band to which we have 
access from the Villa de Leyva site. 
This band covers the sky region 
$0^h < \delta < 24^h$, 
$-24^{\circ} \ 22^{\prime} < \alpha < +35^{\circ} \ 37^{\prime}$
in right ascension and declination respectively.  
Figure \ref{FigGEM2CA} is a rendering of this map in
an equatorial celestial projection and Figure \ref{FigGEM2CB} represents
the same data in contour levels (Kelvin). 
The number of observations per pixel goes from 5 to 
1147 with a mean of 206 and a sigma of 137.
The range of values of the map temperature is 0 to 4.4 K,
but large systematic errors,e specially in calibration, need to be 
taken into account. 

\section{Frequency Dependence}
The spectral index of Galactic synchrotron emission can be readily determined 
at frequencies less than 1 GHz where the observational baselevel uncertainty is 
much less than the total Galactic emission.  
Lawson et al. (1987) used data covering the range 38 to 1420 MHz to determine 
the spectral index variation over the northern sky.  
Clear variations in spectral index of at least 0.3 were found.  
A steepening in the spectral index at higher frequencies in the brighter 
features such as the loops and some SNRs.  
Up to 1420 MHz no such steepening was found in the regions of weaker emission.  
At the higher Galactic latitudes where no reliable zero level is available 
at 1420 MHz, an estimate can be made of the spectral index of local features 
by using the T-T technique.  
The de-striped 408 and 1420 MHz maps gave temperature spectral indices 
of 2.8 to 3.2 in the northern galactic pole regions 
(Davies, Watson, Gutierrez 1996).

At frequencies higher than 2-3 GHz no reliable zero levels have been measured 
for large area surveys.  
The relevant observational material comes from beam switching 
or interferometric data.  
Accordingly the spectrum of individual features is estimated.  
On scales of a few degrees the temperature spectral index between 5 GHz 
(Melhuish 1997) and 408 MHz was $\sim 3.0$.  
Similarly the 10, 15, 33 GHz beam switching data (Hancock et al. 1997)
also indicated a spectral index of $\sim 3.0$ at 10 GHz.  
These results show that at higher Galactic latitudes synchrotron emission 
dominates up to 5 GHz and most likely to 10 GHz.

Platania et al. (1997) has used high frequency data from a number 
of sky locations to investigate the frequency dependence of the 
synchrotron emission spectral index and also find evidence for a steepening.
Based on the local cosmic ray electron energy spectrum, 
the synchrotron spectrum should steepen with frequency 
to about -3.1 at these higher microwave frequencies.
In addition their comparison of a 19 GHz map (Cottingham 1987) and the 408 MHz
map (Haslam et al. 1982) showed clear evidence for spatial variation
in the synchrotron spectral index.
The North Polar Spur is quite evident as its synchrotron spectral index
is steeper than the average as was previously known.
A part from this feature, the synchrotron spectrum appears
flatter than average at high Galactic latitudes.

\section{Separation of Signals}
In order to interpret observations it is necessary to separate the
various Galactic emissions from the CMB anisotropies 
and extragalactic foregrounds.
The CMB, thermal, and non-thermal emissions can be separated by virtue 
of their differing frequency spectral indices.
Free-free emission in the optically thin domain, 
appropriate to CMB observations at 30 GHz and above, 
has a brightness temperature spectral index of 2.1 $\pm$ 0.1.  
In the case of synchrotron emission from the Galaxy the spectral index 
at frequencies up to a few GHz is 2.6 to 3.0 (Lawson et al. 1987).  
The synchrotron spectral index varies from position to position 
whether it is measured in SNRs 
or in the more extended Galactic features.  
At a given position the synchrotron 
spectral index is expected to change with frequency. 
Optically thick synchrotron emission gives a strongly inverted spectrum 
($S_\nu \propto \nu^{5/2}$), independent of the energy distribution 
of electrons, producing a low frequency cutoff. For very compact, high 
density regions cutoff frequencies as high as several GHz are observed. 
likewise, compact and dense HII 
regions may be optically thick, and have an inverted spectrum 
($S_\nu \propto \nu^{2}$) up to relatively high frequencies. 
It will be necessary to investigate the possible existence of 
sources with high absorption cutoffs, which 
are undetectable at low frequencies. 
Due to radiation energy losses by the relativistic electrons the 
synchrotron spectrum 
will steepen by 0.5 at higher frequencies;  
the turnover frequency will depend on the age of the electrons and 
the strength of the magnetic field.

The spatial variation of the emissions is also significant
and CMB observations routines discard the data in regions
of high emission, e.g. near and on the Galactic plane. 
Another feature that might be used is the properties of the angular 
power spectrum.
An important consideration in this context is what is 
the angular power spectrum of the synchrotron emission particularly 
at intermediate and high galactic latitudes.  
Before the power spectrum can be determined, it is necessary 
to remove the baseline stripes in the most commonly used radio maps 
at 408 and 1420 MHz.  
These stripes which contain power on angular scales of a few to ten degrees 
can be largely removed (Davies, Watson \& Gutierrez 1996).  
Lasenby (1996) has used the 408 and 1420 MHz surveys to estimate 
the spatial power spectrum of the high latitude region surveyed 
in the Tenerife experiments and found 
an angular power spectrum slightly flatter than $\ell^{-2}$.
G\'{o}rski et al. 1996 found an angular power spectrum $\propto \ell^{-3}$
which is one power steeper than scale-invariant CMB anisotropies.

A complicating issue is that 
synchrotron emission is linearly polarized by its very nature.
Surveys at frequencies less than about 1 GHz where Faraday 
depolarization is significant show polarized structure at a
level of a few tens of percent (e.g. Spoelstra 1984).  
At the higher frequencies sampled by CMB observations Faraday depolarization 
will be minimal and higher percentage polarizations will be expected,
perhaps reaching 30 to 50 percent as found in some SNRs and Galactic spurs.  
Spatial structure in polarization ranges from 10 arcmin to
tens of degrees in the spurs and loops 
(Spoelstra 1984, Wieringa et al. 1993, Duncan et al. 1997).

A number of authors have proposed methods for separating the various signals.
The first systematic approach was to use the frequency dependence
(e.g. Brandt et al. 1994, Dodelson 1995) 
by employing maps at multiple frequencies.
This is not a powerful use of the full information of the observations
and new approaches were proposed including
marginlizing over the extraneous signals (Scott et al. 1996),
Wiener filtering (e.g. Tegmark \& Efstathiou 1996,
Bouchet, Gispert \& Puget 1996, Bouchet et al 1997)
and maximum entropy technique (Hobsen et al. 1998).

\vspace{12.0mm}    

{\bf Acknowledgments.} --- 
This work supported in part by the DOE
contract No. DE-AC03-76SF00098
through the Lawrence Berkeley National Laboratory
and NASA Long Term Space Astrophysics Grant No. \#NAG5-6552.

\end{document}